%% file: OT_paper.tex
\pdfoutput=1
\documentclass[%
 reprint,
 amsmath,amssymb,
 aps,
]{revtex4-1}

\usepackage{graphicx}
\usepackage{dcolumn}
\usepackage{bm}
\usepackage[colorlinks=true,citecolor=blue, urlcolor=blue]{hyperref}
\usepackage{color}
\usepackage{dsfont}

\begin{document}

\preprint{APS/123-QED}
\title{Multidimensional optical trapping of a mirror}

\author{Antonio Perreca}

 \email{aperreca@syr.edu}
 \author{James Lough}
 \email{jdlough@syr.edu}
 \author{David Kelley}
 \email{dbkelley@syr.edu} 
\author{Stefan W. Ballmer}
 \email{sballmer@syr.edu}

\affiliation{%
 Department of Physics, Syracuse University, 
 Syracuse, New York, 13244-1130, USA }



\date{\today}

\begin{abstract}
Alignment control in gravitational-wave detectors has consistently proven to be a difficult problem due to the stringent noise contamination requirement for the gravitational wave readout and the radiation-pressure-induced angular instability in Fabry-Perot cavities (Sidles-Sigg instability).
We present the analysis of a dual-carrier control scheme that uses radiation pressure to control a suspended mirror,
trapping it in the longitudinal degree of freedom and one angular degree of freedom. We show that this scheme can control
the Sidles-Sigg angular instability. Its limiting fundamental noise source is the quantum radiation pressure noise, providing an advantage compared to the conventional angular control schemes.
In the Appendix we also derive an exact expression for the optical spring constant used in the control scheme.
\begin{description}
\vspace{1.2cm}
\item[PACS numbers]
 04.80.Nn, 07.60.Ly, 95.55.Ym
\end{description}
\end{abstract}

\pacs{Valid PACS appear here}
\maketitle

\newcommand{\tcr}{\textcolor{red}}
\newcommand{\tcb}{\textcolor{blue}}
\newcommand{\tcm}{\textcolor{magenta}}
\newcommand{\tcg}{\textcolor{green}}
\newcommand{\tcp}{\textcolor{purple}}
\newcommand{\irm}{\mathrm{i}}


\section{Introduction}
\label{sec:int}

The Laser Interferometer Gravitational-Wave Observatory (LIGO) is part of a worldwide 
effort to detect gravitational waves and use them to study the Universe \cite{BPAbbott09}. Construction of 
LIGO's advanced detectors is underway. The installation is expected to finish in 2014. The goal of Advanced LIGO (aLIGO) is the first direct detection of gravitational waves 
from astrophysical sources such as coalescing compact binaries and core-collapse supernovae.
These detections will open a new spectrum for observing the Universe and establish the field of 
gravitational-wave astronomy. 
These initial observations will also show the potential science gain of further increasing the state-of-the-art sensitivity of gravitational-wave detectors \cite{Smith09,Harry10,Losurdo12}. Such detectors operate near the standard quantum limit, meaning that the contributions from quantum radiation pressure and shot noise are about equal in the observation band \cite{Caves80, Ni86}.

To design a successor to aLIGO, techniques to operate gravitational-wave interferometers below 
the standard quantum limit need to be developed \cite{Dan12, Chen13}. Dual carrier control systems and angular control 
using stable optical springs are promising methods for evading quantum-mechanical limitations on 
detector sensitivity \cite{LIGO10, Braginsky02b, Arcizet06b, Corbitt06b, Kippenberg05, Sheard04}. 
In 2007 Corbitt \emph{et al}. at the LIGO Laboratory at the Massachusetts Institute of Technology 
demonstrated a one-dimensional optical trap of a one gram mirror using a novel two-carrier scheme \cite{Corbitt07}. 
Their work 
clearly demonstrated the potential of this technique. Extended to angular degrees of freedom, it has 
the prospect of opening a completely new approach to the angular control problem in future generation 
gravitational-wave detectors \cite{Punturo10}. 
Sidles and Sigg have shown that, for a Fabry-Perot cavity with a single 
resonating laser field, the radiation pressure force will couple the two end mirrors, always creating one 
soft (unstable) and one hard (stable) mode \cite{Sidles06}. This sets a lower limit on the required angular control 
bandwidth, which inevitably results in higher noise contamination by angular control noise and limits the angular control performance in the first and second generation 
gravitational-wave interferometers \cite{LIGO10, Braginsky01, Dooley13, Hirose10}. 
As we will show in Sec. \ref{sec:IV}, angular optical trapping can bypass the Sidles-Sigg instability. Its fundamental noise limit is quantum radiation pressure noise. By design it is not affected by sensing noise, making it a promising candidate for low-noise angular control.
Additionally, 
optical trapping can be used to cool a mechanical degree of freedom. Radiation pressure-based cooling is the preferred approach for cooling to the quantum ground state in the limit where the cavity line width is smaller than the mechanical frequency  (good cavity limit)\cite{Genes08}. It can enable the manipulation of a macroscopic object at the quantum level \cite{Teufel11, OConnell10, Chan11, TCorbitt07, Matsumoto13}. 
However reaching the quantum ground state requires reducing the total rms motion, rather than the spectral density in the frequency band above the mechanical suspension resonance, as desired for a low-noise angular control system. We therefore will not further explore reaching the quantum ground state.

In this paper we present
a prototype of a position and yaw optical trap for a suspended test mirror using a double dual-carrier control 
scheme. 
With mechanical suspension frequencies around $1\,{\rm Hz}$ such a system is, in virtually all cases, in the bad cavity limit; i.e., the cavity line widths are larger than the mechanical frequencies.
We propose a system with two longitudinal traps acting on different spots of a single mirror; together, these traps will constrain both the position degree of freedom and one angular degree of freedom of the mirror.
This essentially replaces the current magnetic drives with optical traps.  The idea is promising and will be easy to apply to the other
angular degree of freedom.
The model includes two optical cavities with the trapped end-mirror in common. Each cavity is illuminated with two overlapping laser beams at different frequency detunings: one is positive detuned (blue detuning) and the other is negative detuned (red detuning).
The two dual beams form two statically and dynamically stable optical springs with different lever arms and different power, designed such that the static (commonly named DC) radiation pressure torques of the two dual beams cancel each other while 
DC radiation pressure force is canceled by displacing the position pendulum. 

As a result, by picking the right 
parameters, we can obtain a system that is stable in the longitudinal and angular degrees of freedom with a mirror 
displacement range of the order of picometers.

The outline of this paper is as follows. In Sec. \ref{sec:II} we review the idea of an optical spring. We then couple optical springs to a mechanical system and analyze the stability of the resulting optomechanical system.
Section \ref{sec:III} extends the stability analysis to more than one dimension.
In Sec. \ref{sec:IV} we show that such a two-dimensional optical spring is necessarily stronger than the Sidles-Sigg instability. In Sec. \ref{sec:V} we calculate the radiation pressure noise, which is the fundamental limiting noise for radiation pressure control. Finally, in Appendix \ref{app:A}, we derive the approximation-free expression for the optical spring in a Fabry-Perot cavity, which to our knowledge has not been published yet.


\section{Stability principle}
\label{sec:II}

An optically detuned Fabry-Perot cavity naturally leads to a linear coupling between intracavity power and 
mirror position. Depending on the sign of the detuning, this coupling creates an optical spring which
is either statically stable or unstable. Due to the time delay in the optical field build-up, the optical spring 
restoration force is slightly delayed. This leads to a dynamically unstable spring for the statically stable case
and a dynamically stable spring for the statically unstable case. Corbitt \emph{et al}. \cite{Corbitt07} demonstrated that by adding a second, frequency-shifted optical field (subcarrier) with a different detuning and power, a statically and dynamically stable optical spring can be achieved. The dual-carrier scheme has been used to optically trap a gram-scale mirror, controlling its longitudinal degree of freedom.
Moreover, the damping of the optical spring can be controlled by adjusting the detuning of both carrier and subcarrier and their relative amplitudes. This naturally allows for efficient cooling of the degree of freedom seen by the optical spring. In contrast to a mechanical spring, this damping does not introduce intrinsic losses, and thus does not contribute to the thermal noise.

This technique can be extended to alignment degrees of freedom. By duplicating the Corbitt \emph{et al}. approach for trapping 
with a second, different, optical axis and a different beam spot on the controlled mirror, it is possible to control the angular 
degree of freedom with radiation pressure alone.

To be able to understand the stability of multidimensional optomechanical systems, we first recall the simple driven damped mechanical oscillator. From there we will stepwise increase the complexity by adding optical springs and additional degrees of freedom. 

\subsection{Damped mechanical oscillator stability}

Although the damped mechanical oscillator is a well known system, we will take it as a starting point to make the reading clearer. Our goal is to describe the mechanical oscillator in the language of control theory, which allows us to understand the stability of the system from a different point of view. This approach can then be naturally extended to include the effect of additional optical springs. 

The motion of a harmonic oscillator of mass $m$, spring constant $k_m$ and velocity damping $b$, driven by the external force $F_{ext}$, can be expressed as \cite{Saulson90}
\begin{eqnarray}
\label{eqn:motion}
m\ddot{x}=-k_m x-b\dot{x}+F_{ext}
\end{eqnarray}
where $b$ is also called the viscosity coefficient. Often the damping rate $\Gamma=b/(2 m)$ is used instead.
Traditionally the Eq. (\ref{eqn:motion}) is directly used to get the system's position response $x$ when applying the external force $F_{ext}$. The resulting transfer function is
\begin{eqnarray}
\label{eqn:TF}
G=\frac{x}{F_{ext}}=\frac{1}{-m\Omega^2+k_m+ib\Omega}                                                 
\end{eqnarray}
where $\Omega$ is the angular frequency of the motion.

Alternatively we can describe a damped mechanical oscillator as a feedback system,  with the plant being just a free test mass described by the transfer function 
$M=x/F_{ext}=-1/m\Omega^2$,
obtained directly from the equation of motion of a free test mass. 
The control filter of the feedback loop is the mechanical spring, which takes the mass displacement $x$ as input and acts on the plant with the control signal, or force, $F_K$, which is subtracted from the external force $F_{ext}$.
The transfer function of the control filter is $K_M=F_{K}/x=k_m+ib\Omega$. In this picture we can now calculate the closed loop transfer function and obtain the same expression as in Eq. (\ref{eqn:TF}),
\begin{eqnarray}
\label{eqn:TF_fm}
G=\frac{M}{1+K_M M}=
\frac{1}{-m\Omega^2+k_m+ib\Omega}
\end{eqnarray}
where $OL_M=-K_M  M = (k_m+ib\Omega)/m\Omega^2$ describes the open loop transfer function of the system.


\subsubsection{Stability}
We can now check for the stability of the system in both pictures.
We recall from literature that the stability of a system described by its transfer function $G$ can be evaluated looking at the poles 
of its transfer function in the s-plane ($s=i\Omega$) \cite{Greensite70}. In particular a system is stable only if its transfer function's poles have
a negative real part,  and the multiplicity of poles on the imaginary axis is at most 1.
The transfer function in Eq. (\ref{eqn:TF}) has the following poles:
\begin{eqnarray}
\label{eqn:poles}
i\Omega=-\frac{b}{2m}\pm\sqrt{\frac{b^2}{4m^2}-\omega_0^2},
\end{eqnarray}
where $\omega_0^2=k_m/m$ is the resonant frequency of the pendulum. 
The value of the damping rate $\Gamma=b/2m$ compared to $\omega_0$ determines whether the system is overdamped, underdamped or critically-damped. But since  $\Gamma$ (or $b$) is always positive, 
the real part of the poles is always negative. The system is thus always stable. 

From the control theory point of view, the stability can also be evaluated with no loss of generality by considering the open loop transfer function $OL_M= (k_m+ib\Omega)/m\Omega^2$ and applying, for example, the Bode stability criterion \cite{Franklin94}. The positivity of $b$ guarantees an always positive phase margin and therefore stability.
In the reminder of this work, for simplicity, we will test the stability of the control scheme using the Bode graphical method.


\subsection{Optical spring: A classical model}
Next, we look at an optical spring.
We start with a Fabry-Perot  cavity of length 
$L_0$, frequency detuning $\delta$ (rad/Hz), amplitude transmittance coefficients $t_1$, $t_2$  and amplitude reflectance coefficients $r_1$, $r_2$ of the input and output cavity mirror respectively. 
The light field inside the cavity builds up and exerts a radiation pressure force on both mirrors.

We define the propagator $X=r_1r_2e^{-2i\delta\tau}$ and phase factor $Y=e^{-i\Omega\tau}$, with $\tau=L_0/c$ the one-way
travel time of the photon inside the cavity, $k$ is the wave vector of the light field  and $\Omega$ 
is the mechanical frequency of the pendulum. From this we can obtain an elastic force-law for small displacement values $x$, but potentially large detuning from resonance:
\begin{eqnarray}
\label{eqn:Frd}
F_{rad}=F_0-K_{OS}\cdot x + O(x^2),
\end{eqnarray}
where
\begin{eqnarray}
\label{KOS_full_2}
K_{OS}=K_0\left [ \frac{Y^2}{(1-Y^2X)(1-Y^2\overline{X})}  \right ]
\end{eqnarray}
is the optical spring constant and $\overline{X}$ is the complex conjugate of $X$. Here $K_0$ is the 
(mechanical) frequency-independent part of the spring constant:
\begin{eqnarray}
\label{eqn:K0}
K_0=F_0 \cdot 2 i k \cdot (X-\overline{X}),   \quad \mbox{with}\nonumber\\ 
F_0 = P_0 \cdot \frac{2  r_2^2}{c} \cdot \frac{t_1^2}{(1-X)(1-\overline{X})}
\end{eqnarray}
The expression in Eqs. (\ref{KOS_full_2}) and (\ref{eqn:K0})
is the general expression for $K_{OS}$ up to linear order in $x$. While approximations for this formula have been published before \cite{Barginsky02}, we are not aware of a previous publication providing the full expression.
We address the complete derivation of the optical spring constant $K_{OS}$ in Appendix \ref{app:A}. There we also show that with the approximations $2\Omega\tau\ll1$ and $2\delta\tau\ll1$  Eq. (\ref{KOS_full_2}) is equivalent to the expressions already existing in literature \cite{Barginsky02,Corbitt07}. 

We note that $K_0$ is a real number. Its sign is determined by the imaginary part of $X$. A positive sign is associated with positive detuning ($\delta>0$) and a restoring force (statically stable),  while a negative sign is due to  negative detuning ($\delta<0$) and
leads to a antirestoring force  (statically unstable).  Also, for small (positive) frequencies $\Omega\tau\ll1$, the sign of the imaginary part of Eq. (\ref{KOS_full_2}) is opposite to its real part, leading to positive dynamic feedback for the statically stable case and  negative dynamic feedback for the statically unstable case.

\begin{figure}[htbp]
	\centering
		\includegraphics[width=8cm]{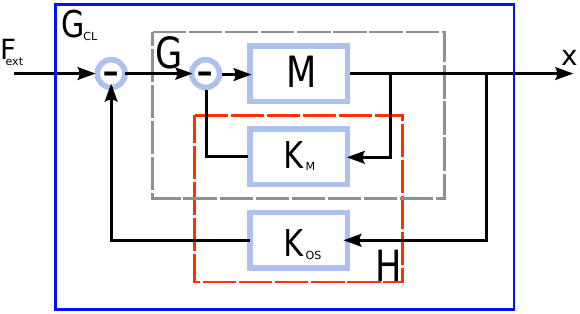}
	\caption{{Mechanical oscillator and feedback systems. The mechanical oscillator can be seen as plant ($G$) and the optical spring $K_{OS}$ as feedback or
	alternatively as free test mass (plant $M$) and $H=K_{OS}+K_M$ as feedback. 
	Both the cases lead to the same closed loop transfer function $G_{CL}$ which describes the system as a damped mechanical oscillator in the presence of
	the optical spring, which is subjected to the external force $F_{ext}$ and has the corresponding displacement $x$ as output.}}
	\label{fig:blocks2}
\end{figure}

Our next step is to couple the optical spring to a mechanical pendulum. We can treat this as either a damped mechanical oscillator with transfer function $G$, controlled by an optical spring $K_{OS}$, or as a free mass with transfer function $M$, controlled by the total feedback filter $H = K_M + K_{OS}$, see Fig.\,\ref{fig:blocks2}.
In both cases we obtain the same closed-loop transfer function, equivalent to the one we would have obtained by
rewriting the equation of motion of a damped mechanical oscillator with an optical spring:
\begin{eqnarray}
\label{eqn:TFco}
G_{CL}=\frac{x}{F_{ext}}=\frac{G}{1+K_{OS} G}=\frac{M}{1+H M}\nonumber\\
=\frac{1}{-m\Omega^2+K_M+K_{OS}}
\end{eqnarray}

The stability of the total system can again be evaluated  by either looking at the poles of the closed-loop
transfer function $G_{CL}$, or looking at the gain and phase margin of the open loop transfer function $OL_{MH}=-H/m\Omega^2$. The latter is generally more convenient. Unless compensated by large mechanical dissipation in $K_M$, the positive dynamic feedback for the statically stable case ($\delta>0$) leads to a dynamically unstable system. 
Intuitively this can be understood as a phase delay in the radiation pressure build-up which is caused by the cavity storage time.
For $\delta<0$ the system is statically unstable.

\subsection{Double carrier spring}

The seemingly intrinsic instability of optical springs can be overcome by a scheme 
proposed by Corbitt \emph{et al}. \cite{Corbitt07}. The carrier is set at a large positive detuning ($\delta>0$, large $|\delta|/\gamma$, where $\gamma$ is the line width). This provides a static restoring force, together with a relatively small dynamic instability (antidamping). Then a subcarrier is added at lower power and with a small negative detuning ($\delta<0$, small $|\delta|/\gamma$). The subcarrier adds sufficient damping to stabilize the total optical spring, while leaving the sign of the static restoring force unchanged.
For appropriately chosen parameters of carrier ($c$) and subcarrier ($sc$) (power $P_0^c$ and $P_0^{sc}$, detuning  $\delta_c$ and $\delta_{sc}$) the resulting total system thus becomes stable.

The spring constant of the total optical spring is simply the sum of the individual spring constants of the carrier and subcarrier
\begin{eqnarray}
\label{eqn:KOSsum}
K_{OS}=K_{OS}^c+K_{OS}^{sc}
\end{eqnarray}
where the individual springs $K_{OS}^c$ and $K_{OS}^{sc}$ are given by Eq. (\ref{eqn:TFco}).


Conceptually we can think of the dual-carrier optical spring as a physical implementation of a feedback control filter for the mechanical system. With this tool at hand, we can start to analyze the behavior and stability of higher-dimensional mechanical systems in the next section.



\section{Control model of longitudinal and angular degrees of freedom}
\label{sec:III} 

\begin{figure}[t]
	\centering
		\includegraphics[width=8cm]{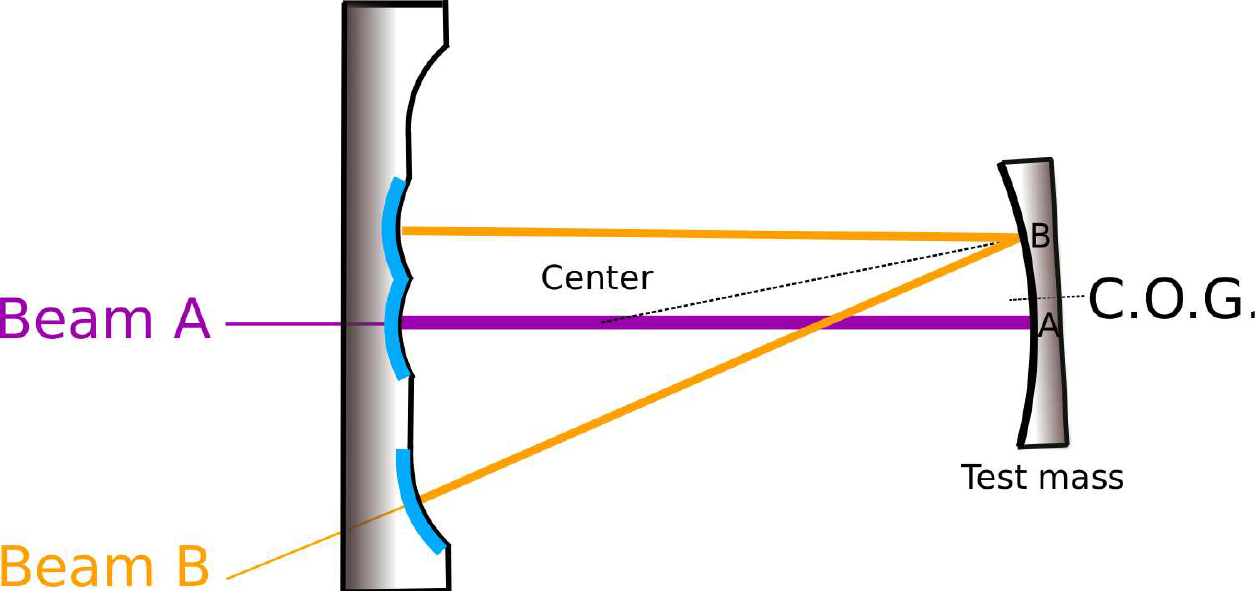}
	\caption{
	In this sketch the main purple (beam $A$) optical axis hits the test mirror 
	at point $A$, slightly displaced from the center of gravity (C.O.G.), such
	 that it still corresponds mainly to the length degree of freedom. Thus the second orange (beam $B$) optical axis, which hits the test mirror closer to the edge at point $B$, needs much less power to balance the total DC torque. In our test setup the large input coupler is a composite mirror. It is 600 times more massive than the small mirror. The choice of a V-shaped beam $B$ results in a more practical spot separation on the input coupler. }	

	\label{fig:angular}
\end{figure}

We will now extend our analysis to additional degrees of freedom. Experimentally, a torsion pendulum suspension is  easy to build. Therefore we will focus our attention to controlling the yaw motion of a test mirror, keeping in mind that the method can be applied to any additional degree of freedom. For actively controlling two degrees of freedom (length and yaw), we need a two-dimensional control system. In other words, we will need a second dual-carrier optical spring in a setup that for example looks like Fig.\,\ref{fig:angular}. We will label the two dual-carrier optical fields as beams $A$ and $B$. Each beam includes a carrier and a subcarrier field, i.e.
\begin{eqnarray}
\label{eqn:beams}
\mbox{beam $A$ = carrier $A$ + subcarrier $A$}\\ \nonumber
\mbox{beam $B$ = carrier $B$ + subcarrier $B$}\nonumber
\end{eqnarray}
The two beams have a different optical axis, and each has its own optical spring constant, $K_{OS}^A$ and $K_{OS}^B$, given by Eq. (\ref{eqn:KOSsum}).

If we define $x_A$ and $x_B$ as the longitudinal displacement of the mirror at the contact points
of beam $A$ and beam $B$ on the test mirror,
 and $F_A$ and $F_B$ as the corresponding exerted forces, we can describe the mechanical system with a plant matrix $M$:
\begin{equation}
 \begin{pmatrix}
x_A\\ x_B
\end{pmatrix} 
=
M \begin{pmatrix}
F_{A}\\ F_{B}
\end{pmatrix}
\label{eq:MF}
\end{equation}
The explicit expression for $M$ for a torsion pendulum is given in Appendix \ref{app:B}.

\begin{figure}[htbp]
		\includegraphics[width=6.1cm]{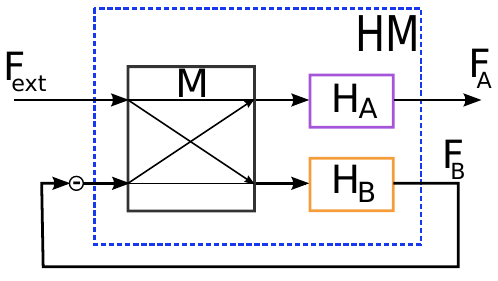}
	\caption{Block diagram of beam $A$ and beam $B$. The transfer function $F_A/F_{ext}$ is equal to $OL_A$ from Eq. (\ref{eq:2dol}). Each loop affects the other resulting in cross terms
	present in the matrix $HM$. $M$ and $H_{A,B}$ are the transfer functions of the mechanical system and the optical springs of beam $A$ and $B$, respectively.}
	\label{fig:block_loops}
\end{figure}

The control is provided by the optical springs. In the $x_A$-$x_B$ basis the control matrix $H$ is diagonal and given by  (also see Fig.\,\ref{fig:block_loops})
\begin{equation}
\begin{pmatrix}
F_{A}\\ F_{B}
\end{pmatrix}
= H
 \begin{pmatrix}
x_A\\ x_B
\end{pmatrix} 
=  \begin{pmatrix}
K_{OS}^A & 0 \\ 0 & K_{OS}^B
\end{pmatrix} 
 \begin{pmatrix}
x_A\\ x_B
\end{pmatrix} 
\label{eq:HX}
\end{equation}

For a multidimensional feedback system to be stable, it is sufficient that each individual (one-dimensional) feedback loop is stable, assuming all remaining control loops are closed. In other words, in our two-dimensional optomechanical system, we close the beam $B$ control filter for evaluating the open loop transfer functions $OL_{A}$, and vice versa. For the open loop transfer functions $OL_{A}$ and $OL_{B}$ we then find: 
\begin{eqnarray}
\label{eq:2dol}
OL_{A}=e_A^{T}\left(\mathds{1}+HM (\mathds{1} - e_A e_A^T) \right)^{-1}HMe_A  \\
OL_{B}=e_B^{T}\left(\mathds{1}+HM (\mathds{1} - e_B e_B^T) \right)^{-1}HMe_B \nonumber
\end{eqnarray}
with $e_A^T=(1,0)$ and $e_B^T=(0,1)$. The derivation of this expression is given in Appendix \ref{app:C}.


\subsection{An example}

It is worth considering a specific set of possible values for our model and evaluate the control of  angular and longitudinal degrees of freedom of a gram-scale test mirror using the radiation pressure of the light.
All the optical fields involved in our analysis are derived from the same wavelength light source through frequency shifting.
The model includes two optical cavities (Fig.\,\ref{fig:angular}), referred to as beam $A$ and $B$, both with an optical finesse of  about $8500$, line width $\gamma/(2 \pi) = 125\,$kHz and mechanical frequency of $\backsim 1\,$Hz. 
The main cavity (beam $A$) is pumped with $1\,$W of carrier light, detuned by $\delta/(2 \pi)= 250\,$kHz (blue detuning, $\delta/\gamma = 2$), and $0.2\,$W of subcarrier light, detuned by $\delta/(2 \pi) =62\,$kHz (red detuning, $\delta/\gamma = -0.5$). This produces a statically and dynamically stable optical spring with a lever arm of $0.8\,$mm, measured from the mirror center of gravity (C.O.G.). A second optical spring (beam $B$) is pumped with 6 times less power of carrier light, detuned by $=186\,$kHz (blue detuning, $\delta/\gamma=1.5$), and $40\,$mW of subcarrier light, detuned by $62\,$kHz (red detuning, $\delta/\gamma=-0.5$). This side cavity has a lever arm of $3.3\,$mm on the mirror, such that the DC radiation pressure torques of beam $A$ and $B$ cancel. The DC radiation pressure force can be canceled by displacing the position pendulum.

\begin{figure}[htbp]
	\centering
		\includegraphics[width=8cm]{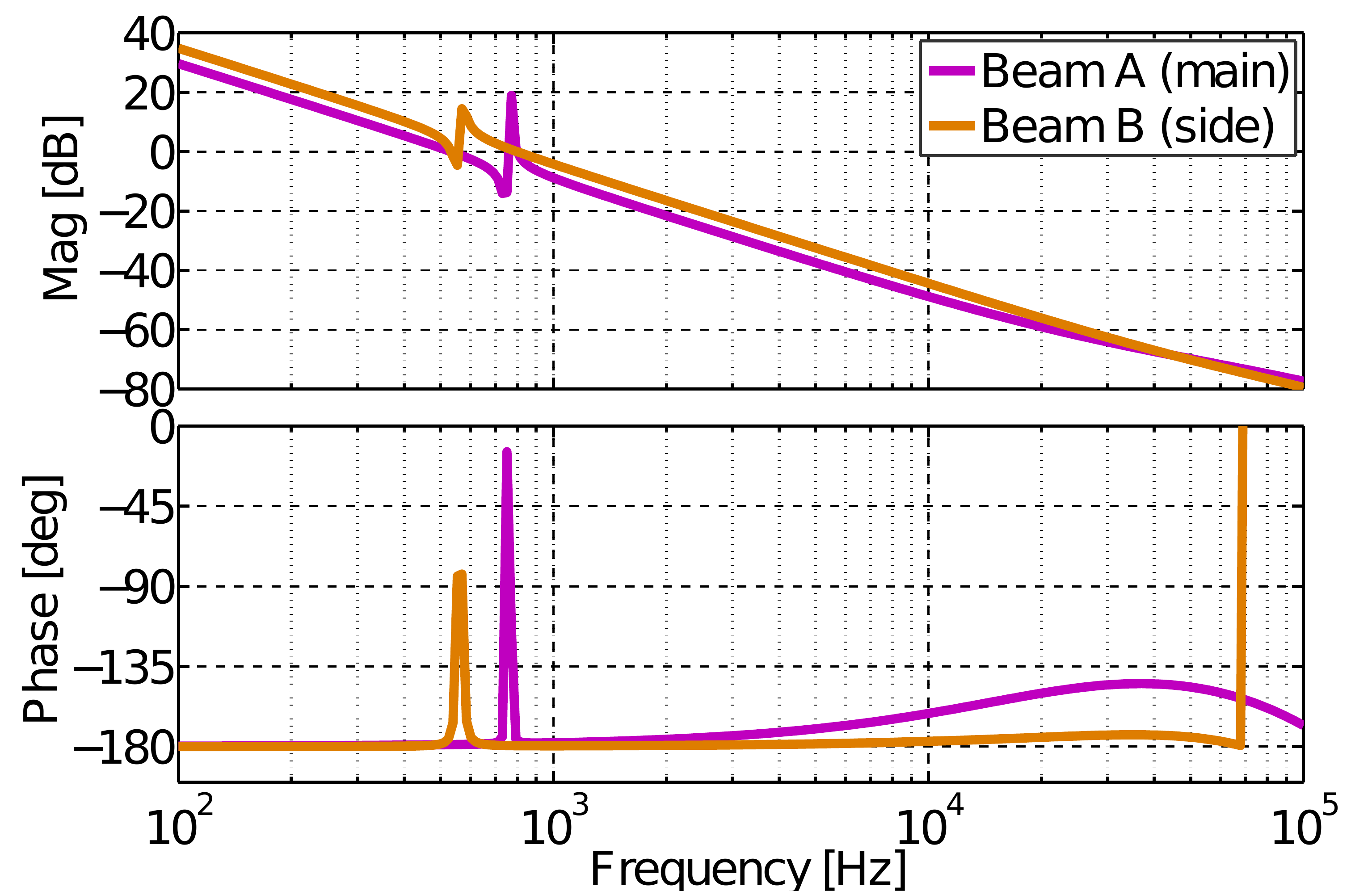}
	\caption{{Open loop gain (OLG) for the main and side cavity.	The respective other loop is closed, and shows up as a resonance in the OLG. Note that, despite multiple unity gain crossings, both loops are stable because the resonances effectively implement a lead filter and the OLG avoids the critical point -1. Thus the dynamic interplay between multiple trapping beams on one payload does not introduce an instability.}}
	\label{fig:control_loops}
\end{figure}

The stability of the combined two-dimensional system is addressed in Fig.\,\ref{fig:control_loops}. Plotted are the open loop gain functions of the two degrees of freedom (the two optical traps) under the assumption that the other loop is closed. The presence of the second loop introduces a resonance feature in each loop at the unity gain frequency of the other loop. However the open loop gain avoids the critical point -1 (phase at zero), leading to a stable system. The model parameters were intentionally tuned for low damping / high quality factor in order to demonstrate that the system remains stable. Lower quality factors, and therefore stronger cooling is easily achievable.


\subsection{Stability range}
\label{sec:stability}
We can now estimate the robustness of our feedback control system 
by changing the microscopic length $x_A$ and $x_B$ of the two cavities. This changes the detuning of the optical springs for both beams. Therefore the propagators $X_A$ and $X_B$ for both beams change according to $X_{A,B}=r_1r_2 e^{-i\delta_{A,B}\tau_{A,B}}\cdot e^{ikx_{A,B}}$. For each position both the static and dynamical stability of the total optical spring system given by Eq. (\ref{eq:2dol}) is reevaluated.

\begin{figure}[htbp]
	\centering
		\includegraphics[width=9cm]{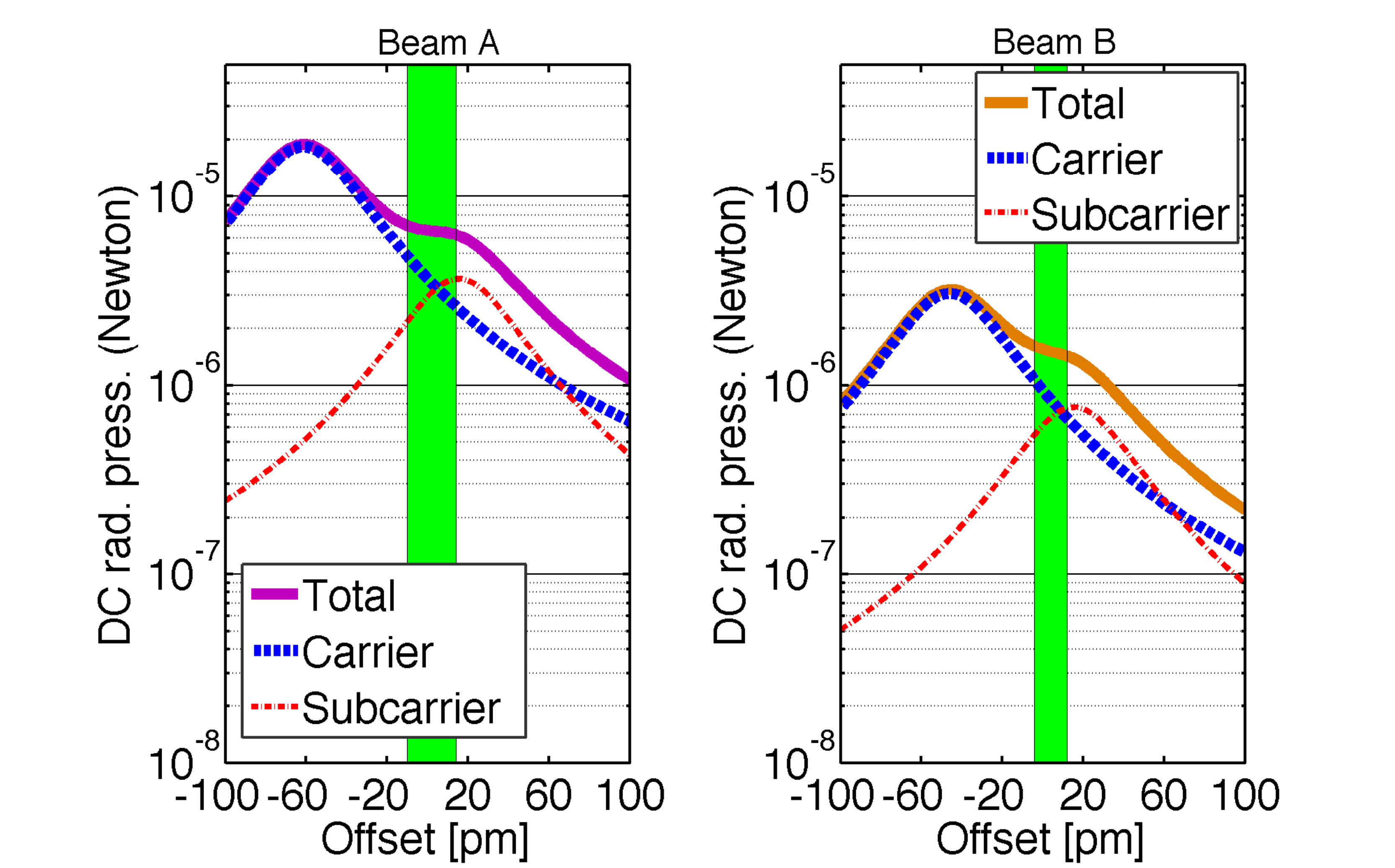}
	\caption{{Static carrier and subcarrier build-up (calibrated in radiation pressure force) as a function of the respective cavity position. Also shown in purple and orange are the total radiation pressure forces of the two cavities. Using the stability testing method from Sec. \ref{sec:stability} we find that the trap is both statically and dynamically stable in the green shaded area.
With the chosen model parameters those regions are about 
20 picometers wide.}}
	\label{fig:stability_region}
\end{figure}

In Fig.\,\ref{fig:stability_region} the radiation pressure force due to the intracavity power of both beams
versus the cavity offset is shown. The green shaded area represents the position range in which the two loops remain stable.  The range is $\backsim 20\,$pm. 
The DC force fluctuations that the system can tolerate are given by the y-axis interval that the total radiation force spends in the green shaded area. 

\section{Angular instability}
\label{sec:IV} 
When operated with high intracavity laser power, suspended Fabry-Perot cavities like the arm cavities of LIGO have a well known angular instability. It  arises from coupling the misalignment of the two cavity mirrors to radiation pressure torques. This is known as the Sidles-Sigg instability \cite{Sidles06}. In this section we show that the intrinsic strength of an optical trap for alignment degrees of freedom is generally bigger, i.e. has a bigger spring constant than any associated Sidles-Sigg instability. 

We start with a cavity of length $L$, with $x_1,x_2$  being the position of the beam spots on mirrors 1 and 2. $\theta_1,\theta_2$ are the yaw angles of the two mirrors and $R_1,R_2$ are their radii of curvature. The corresponding g-factors are $g_{1,2}=1-L/R_{1,2}$.
If one or both of the mirrors are slightly misaligned ($\theta_{1,2}\neq 0$), then the radiation pressure force exerts torques $T_1$ and $T_2$ on the two mirrors, given by (see for instance \cite{Sidles06} or \cite{Ballmer13})
\begin{equation}
\label{SidlesSigg_Basic}
\left(
\begin{array}{c}
T_1\\
T_2
\end{array}
\right)
=
\frac{F_0 L}{1-g_1 g_2}
\left(
\begin{array}{cc}
g_2 & -1\\
-1 & g_1
\end{array}
\right)
\left(
\begin{array}{c}
\theta_1\\
\theta_2
\end{array}
\right),
\end{equation} 
where $F_0=P_0\frac{t_1^2}{(1-X)(1-\overline{X})} \frac{2 r_2^2}{c}$ is the intracavity radiation pressure force. Sidles and Sigg first pointed out that, since the determinant of the matrix in this equation
 is negative, the two eigenvalues have opposite sign. This always leads to one stable and one unstable coupled alignment degree of freedom.

First we note that for a situation in which one mass is sufficiently heavy that we can neglect any radiation pressure effects on it (i.e. $\theta_1=0$), it is sufficient to choose a negative branch cavity (i.e. $g_1<0$ and $g_2<0$) to stabilize the setup. This is for instance the case for the example setup described in Fig.\,\ref{fig:angular}.

Next we want to compare the order of magnitude of this effect to the strength of an angular optical spring. If we call $h$ the typical distance of the beam spot from the center of gravity of the mirror, and $x$ the cavity length change at that spot, the order of magnitude of the optical spring torque is
\begin{eqnarray}
T\approx \frac{F_0 L}{1-g_1g_2}\cdot \frac{x}{h}
\end{eqnarray}
We can express this as the strength of an optical spring located at position $h$. The corresponding spring constant $K_{SS} \approx T/(h x)$. Thus we can see that
\begin{eqnarray}
\label{eqn:KSS_def}
K_{SS} \approx \frac{F_0}{1-g_1g_2}\cdot \frac{L}{h^2}.
\end{eqnarray}
We now consider the adiabatic optical spring ($\Omega=0$) in Eq. (\ref{eqn:K0}).  Expressed in terms of $F_0$, $K_{OS}$ becomes
\begin{eqnarray}
\label{eqn:KOS_exact}
K_{OS}=i F_0 \frac{X-\overline{X}}{(1-X)(1-\overline{X})}   2 k
\end{eqnarray}
Since we operate near the maximum of the optical spring, the order of magnitude of the resonance term can be estimated as
\begin{eqnarray}
\label{eqn:res_est}
\frac{X-\overline{X}}{(1-X)(1-\overline{X})} \approx \frac{-i}{1-|X|}
\end{eqnarray}
Thus we can estimate the magnitude of  $K_{OS}$ as
\begin{eqnarray}
\label{eqn:K0_order}
K_{OS} \approx F_0 \frac{4\pi}{\lambda}\frac{1}{1-|X|} \approx F_0 \frac{4}{\lambda} \mathcal{F}
\end{eqnarray}
where $\mathcal{F}$ is the cavity finesse.
From Eqs. (\ref{eqn:KSS_def}) and (\ref{eqn:K0_order}) we see that the optical spring $K_{OS}$  is much larger than the Sidles-Sigg instability spring $K_{SS}$ if
\begin{eqnarray}
\label{eqn:h2}
h^2 >> \frac{\lambda L}{\pi} \frac{1}{1-g_1 g_2} \frac{\pi}{4 \mathcal{F}}
\end{eqnarray}
Now recall that the beam spot size in a Fabry-Perot cavity is given by \cite{Siegman86}
\begin{equation}
w_1^2 = \frac{\lambda L}{\pi} \sqrt{\frac{g_2}{g_1(1-g_1 g_2)}}
\label{equ:spotsize1}
\end{equation} 
Assuming a symmetric cavity ($g_1=g_2$) for simplicity, we thus find that $K_{OS}$  dominates over $K_{SS}$ if
\begin{eqnarray}
\label{eqn:h2w}
h^2 >> w_{1,2}^2 \frac{1}{\sqrt{1-g_1 g_2}} \frac{\pi}{4 \mathcal{F}}
\end{eqnarray}
This condition is naturally fulfilled since we need to operate the angular optical spring with separate beams ($h>w_{1,2}$) and a large finesse ($\mathcal{F}>>1$). Therefore the angular optical spring is indeed strong enough to stabilize the Sidles-Sigg instability.

\input{sectionV}

\section{Conclusions}
In conclusion, we investigated the use of the radiation pressure of laser light as an alternative to a conventional feedback system for controlling the 
longitudinal and angular degrees of freedom of a mirror.
The method is based on a double dual-carrier scheme, using a total of four detuned laser fields in two cavities. 
The two dual-carrier beams hit the mirror in separate spots, forming two stable optical springs.
This constrains both the longitudinal and the angular degrees of freedom of the mirror, replacing completely the commonly used electronic feedback system.
We showed that this setup allows a stable control of the two degrees of freedom, within a displacement range of the test mirror of $\sim 20\,$pm. This promising idea can be extended to the other angular degree of freedom.
We found that such a method creates an angular optical spring stronger than the angular Sidles-Sigg instability, which drives the requirement for angular control in the high power arm cavities of gravitational-wave detectors. We also showed that the fundamental limit of this scheme is the quantum radiation pressure noise, resulting in a reduction in control noise compared to a conventional active feedback approach. 
We are working towards the experimental demonstration of this effect for a gram-scale mirror and beginning to explore its extension
to large-scale gravitational-wave detectors.


\begin{acknowledgments}
We would like to thank Peter Saulson, Prayush Kumar, Riccardo Penco and Matt West for the many fruitful discussions. This work was supported by the National Science Foundation grant PHY-1068809. This document has been assigned the LIGO Laboratory document number  LIGO-P1300224.
\end{acknowledgments}

\appendix
\input{OT_paper_append_2}

\nocite{*}

\bibliography{OT_paper}
\end{document}

%% file: sectionV.tex
\section{Radiation pressure noise}
\label{sec:V}

Another advantage of radiation pressure angular control, compared to a classical approach based on photo detection and feedback, is its fundamental noise limit. 
The classical approach used in gravitational-wave detectors measures angular displacement of a single beam using wave-front sensors.
Unlike that control method, the shot noise and other sensing noises 
never enter a radiation-pressure-based feedback loop. Even though technical laser noise is typically bigger in the simple cavity setup discussed in this paper, the only fundamental noise source of the scheme is quantum radiation pressure noise. In this section we give the full expression for radiation pressure noise in the case of a dual-carrier stable optical spring.

First, we note that as long as we are interested in frequencies much smaller than the any of the features in the detuned cavity transfer function, the radiation pressure noise is relatively simple. If we also assume that the end mirror has a reflectivity of 1, the one-sided ($f\ge0$) radiation-force amplitude spectral noise density is given by
\begin{eqnarray}
\label{eqn:simpleRPN}
S_F(f) = \frac{2}{c} G_{DC} \sqrt{2 \hbar \omega P_0}
\end{eqnarray}
where $G_{DC}$ is the power gain of a static cavity in the detuned configuration, $P_0$ is the power of the shot noise limited beam entering the cavity, and $\omega$ is its frequency.
Equation (\ref{eqn:simpleRPN}) is valid for carrier and subcarrier separately.
Note that this equation does not hold if the end mirror has a finite transmissivity, as quantum fluctuations entering from that port will also contribute to the intracavity shot noise. In the case of a critically coupled cavity, this will result in an increase of the intracavity radiation-force amplitude spectral noise density by exactly a factor of 2.

\begin{figure}[htbp]
	\centering
		\includegraphics[width=8.5cm]{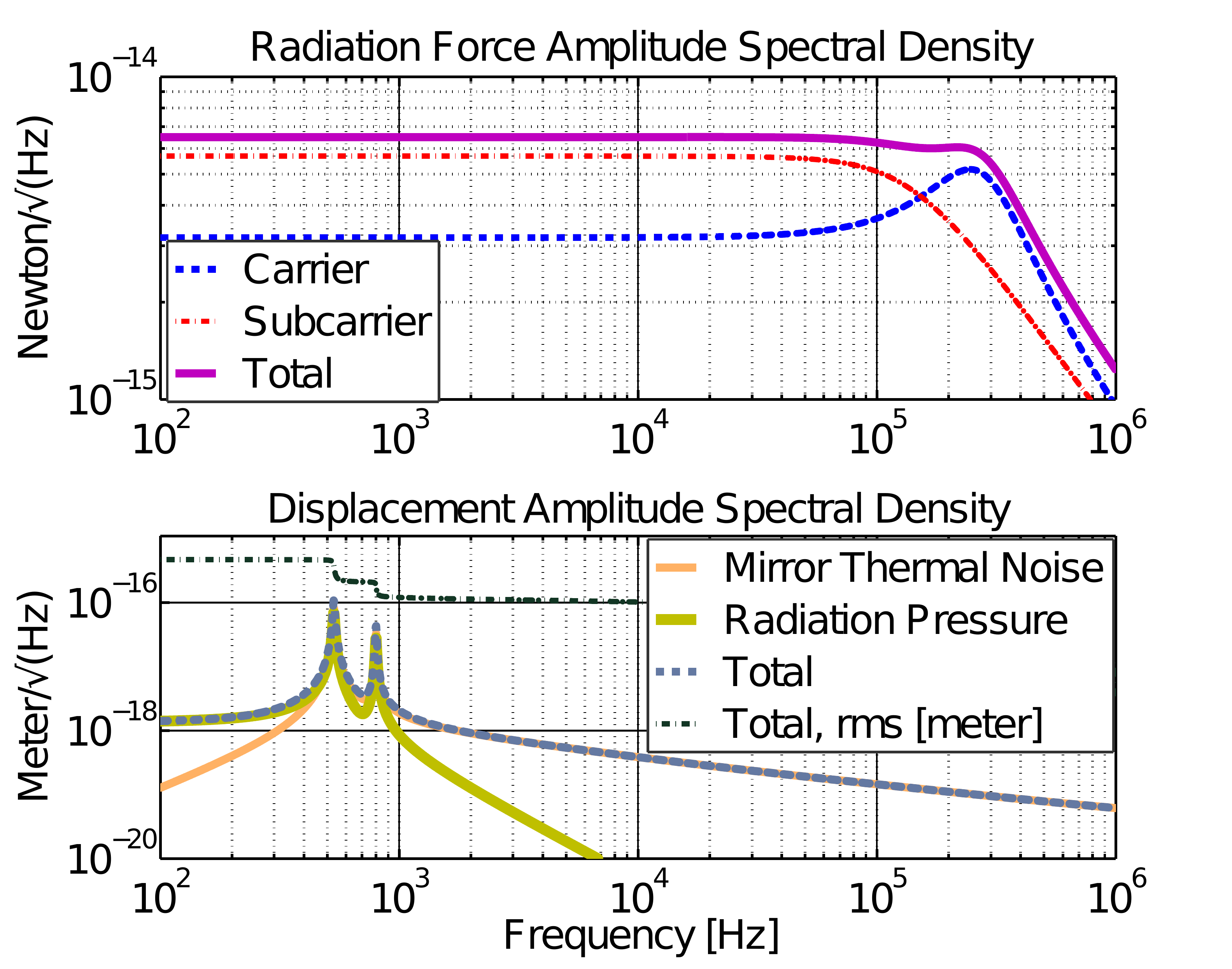}
	\caption{(Top) Radiation force amplitude spectral density for the dual-carrier optical spring used in beam $A$ of the above example. The subcarrier dominates the noise at low frequency, but the higher-power carrier contributes more at high frequencies. Also note that if we choose the same free spectral range for the two carriers, there would be an additional beat note at the difference frequency of $310~{\rm kHz}$. (Bottom)  Radiation pressure and thermal noise displacement amplitude spectral density. The radiation pressure noise is calculated using the optomechanical response given in Eq. (\ref{eqn:closedloop_tf}). The thermal noise is based on a theoretical calculation described in \cite{Saulson90}, \cite{Ballmer13}. Since seismic and suspension thermal noise depend on the experimental implementation, they are not shown, but they would also be suppressed by the optical spring closed loop response. The residual rms motion due to the shown noise sources is less than $10^{-3}$ picometers. With the total rms motion smaller than the 20 picometer stability band shown in Fig.\,\ref{fig:stability_region}, the two cavities will remain locked purely due to the radiation pressure trapping force.}
	\label{fig:RFASD}
\end{figure}

To calculate the exact expression for the radiation pressure noise induced cavity fluctuations, including behavior near the cavity pole frequency, we first realize that we can calculate the radiation-force amplitude spectral noise for a static cavity, and then compute the response of the dual-carrier optical spring system to that driving force. This yields the correct answer up to first order in the size of the quantum fluctuations. For the calculation we track the quantum vacuum fluctuations entering at both ports of the cavity. We introduce $F$, the amplitude build-up factor for a fluctuation at frequency $f = \frac{\Omega + \delta + \omega_{res}}{2 \pi}$:
\begin{eqnarray}
\label{eqn:RPNfunction}
F(f) = \frac{1}{1-XY^2}  =& \frac{1}{1-r_1r_2e^{-2 i \delta\tau}e^{-2 i \Omega\tau}}
\end{eqnarray}
Thus, the total buildup for fluctuations entering through the input coupler (1) and the end mirror (2) are
\begin{eqnarray}
\label{eqn:RPNfunction12}
 t_1 F(f) \,\,\, {\rm and} \,\,\ r_1 t_2 F(f), 
\end{eqnarray}
where we already dropped the one-way propagation factor because it drops out in the radiation force noise calculation below. 
We can now introduce the notation $F_0=F(f_0)$, $F_+=F(f_0+f)$ and $F_-=F(f_0-f)$. We then get the following expression for the one-sided radiation-force power spectral density for either carrier or subcarrier.
\begin{eqnarray}
\label{eqn:RPN_P}
S_F (f) = \frac{2}{c} S_P (f) \,\,\,\, {\rm and} \,\,\,\, S_P(f) = G(f)\sqrt{2\hbar\omega P_0}
\end{eqnarray}
\begin{eqnarray}
\label{eqn:RPN}
G^2(f) = \frac{1}{2}t_1^2|F_0|^2 (t_1^2 \!\!+\! r_1^2t_2^2)( |F_+|^2 \!\!+\!  |F_-|^2) 
\end{eqnarray}
Here $P_0$ is the entering carrier power, and $f_0$ is its frequency. We can see that we recover Eq. (\ref{eqn:simpleRPN}) in the limit $t_2 \rightarrow 0$ and $G/t_1^2=|F_0|^2=|F_+|^2=|F_-|^2$. The resulting force noise from carrier and subcarrier for the cavity A in the example above (see Fig.\,\ref{fig:angular}) is plotted in Fig.\,\ref{fig:RFASD} (top).

Next we calculate the response of the coupled optomechanical system to this driving force, using the following closed loop transfer function obtained from Eqs. (\ref{eq:MF}) and (\ref{eq:HX}):

\begin{eqnarray}
\label{eqn:closedloop_tf}
x = {M}({1+HM})^{-1}F
\end{eqnarray}

Above the optical spring resonances this leads to a $1/f^2$ falloff of the displacement noise, as expected for radiation pressure noise. Meanwhile below the resonance, due to the closed loop suppression, we will have a flat displacement noise. 
Figure\,\ref{fig:RFASD} (bottom) illustrates this in the case of the two-dimensional angular trap discussed above. The level of this flat displacement noise below the unity gain frequency, or optical spring resonance, is at
\begin{eqnarray}
\label{eqn_newraddisp}
S_x(f) =& \frac{S_F(f)}{K_{OS}} \\
\backsim & \frac{\lambda}{\mathcal{F}{P_0}}\sqrt{2 \hbar \omega P_0}
\end{eqnarray}
where we used Eqs. (\ref{eqn:K0_order}) and (\ref{eqn:simpleRPN}) for the estimate, and $\mathcal{F}$ is the cavity finesse.

To compare this noise limit with existing schemes we will consider three angular control schemes: wave front sensing with a single beam (as seen in modern gravitational-wave detectors \cite{Dooley13, Hirose10}), two spatially separated beams with stable optical springs, and an intermediate scheme of two spatially separated beams locked with no detuning using the Pound-Drever-Hall technique \cite{Black01}. 

First we compare the sensitivity to a cavity locked with a Pound-Drever-Hall classical feedback scheme. For the sake of this comparison we want the same dynamics, i.e. the same unity gain frequency and roughly the same loop-shape as in the optical spring system. We can however vary the input power. In addition to radiation pressure noise
we now also have sensing noise.
Photo diode sensing is limited by photo diode quantum efficiency and other factors such as modulation depth, mode matching and overlap. Additionally, in gravitational-wave interferometers the available beam pick-off fraction for alignment sensing is tiny.
All of these factors are typically less than or equal to one, which causes a relative increase in the sensing noise. However, we will not consider these effects for the moment so that we can simply illustrate our point. At best we can use all available power and only have shot noise to worry about. Then the sensing noise is given by
\begin{eqnarray}
\label{eqn:classyShot}
S_x \backsim \frac{\lambda}{\mathcal{F} P_0}\sqrt{2 \hbar \omega P_0}
\end{eqnarray}
We are interested in the noise in the frequency band between the mechanical resonance frequency and the unity gain frequency of the control loop. In this band the radiation pressure noise is loop-suppressed to the level of Eq. (\ref{eqn_newraddisp}), while the displacement noise due to sensing noise is given by Eq. (\ref{eqn:classyShot}). At the nominal power $P_0$ the two schemes are the same. If we now vary $P_0$, we find that the displacement due to sensing noise scales as $P_0^{-1/2}$, while the displacement due to radiation pressure noise scales as $P_0^{1/2}$ [see Eq. (\ref{eqn:simpleRPN})]. Note that we keep the feedback gain in Eq. (\ref{eqn_newraddisp}) equal to the unchanged reference optical spring $K_{OS}$ in order to maintain the same unity gain frequency. We conclude that the lowest total noise, and therefore the best classical feedback scheme, can be achieved at the same power the optical spring operates. Thus the classical scheme can achieve about the same sensitivity as the optical spring system, but in practice performs worse due to real-world sensing limitations.

Finally we want to compare the displacement noise of Eqs. (\ref{eqn:classyShot}) and (\ref{eqn_newraddisp}) to a wave front sensing scheme.
The approximate shot noise limited sensing noise for beam angular and transverse position mismatch, $S_\theta$ and $S_w$, of a wave front sensing scheme is given by
\begin{eqnarray}
\label{eqn:WFSShot}
S_\theta \backsim \frac{\theta_0}{P_0} \sqrt{2 \hbar \omega P_0} \\
S_w \backsim \frac{w_0}{P_0} \sqrt{2 \hbar \omega P_0}
\end{eqnarray}
where the divergence angle $\theta_0$ and waist size $w_0$ of the resonant beam in the cavity are related to the wave length through $\theta_0  w_0  \pi  = \lambda$ \cite{MavThesis}. We can directly compare this wave front sensing scheme to Eq. (\ref{eqn_newraddisp}) if we divide our result by the beam separation $d$.
As long as we choose the beam separation $d$ to be larger than spot size $w$,
 the angular sensitivity of a two-beam system such as the optical spring system
is better than the wave front sensing scheme 
by a factor given by the cavity finesse. Intuitively this result can be understood because having two cavity resonance conditions in the two cavities restricts the angular deviations much tighter than in a one-cavity case. 

%% file: OT_paper_append_2.tex
\section{OPTICAL SPRING CONSTANT DERIVATION}
\label{app:A} 

In this section we consider the effect of light stored in a detuned Fabry-Perot cavity using a classical approach.
The intracavity power generates radiation pressure that exerts on the cavity mirror a force $F_{rad}=-K_{OS}\cdot x$,
where $x$ is the mirror displacement and $K_{OS}$ is the optical spring constant.
Here we show the full derivation of the optical spring constant $K_{OS}$.

We consider a suspended Fabry-Perot cavity of length $L_0$ 
with an incident beam of wavelength $\lambda$ and power $P_0$.
First we calculate a general expression of the intracavity power and then its  radiation pressure force exerted on the end mirror.\\

\begin{figure}[htbp]
	\centering
		\includegraphics[width=8cm]{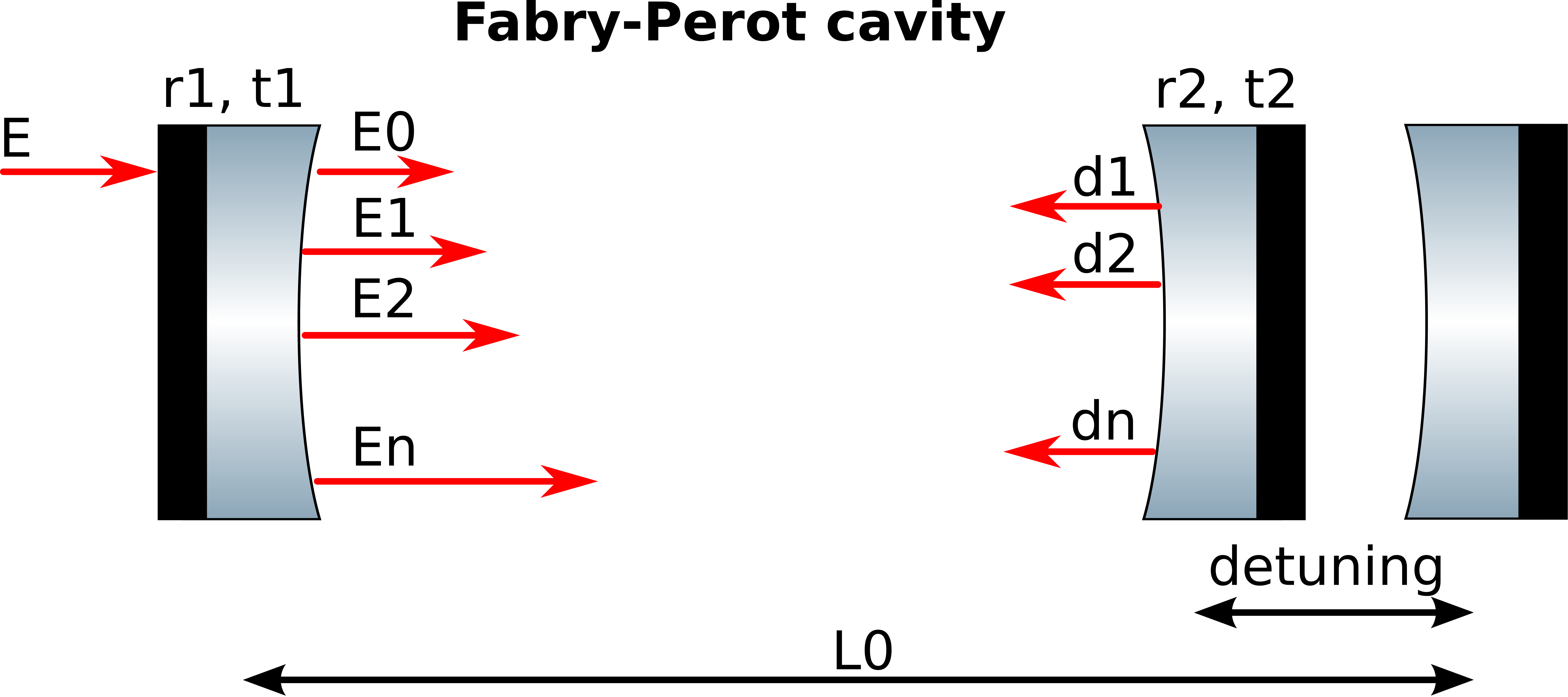}
	\caption{A Fabry-Perot cavity of length $L_0$ and coefficients $r_1,t_1$ and $r_2,t_2$ for the input and end mirrors respectively. 
	The input mirror is stationary while the end mirror is affected by harmonic motion. The incoming field $E$ at each round-trip $i$ adds up a phase shift due to the displacement $d_i$}
	\label{fig:cavity_k}
\end{figure}

The field $E=A_0e^{i\omega t}$ enters the cavity (shown in Fig.\,\ref{fig:cavity_k}) through the input mirror of coefficient $t_1=t$ and $r_1$ and the field inside the cavity at the input mirror can be seen as the following:

\begin{eqnarray}
E_{tot}=E_0+E_1+E_2+E_3+...+En+...
\end{eqnarray}

We consider in our model the following definitions, with $d_n$ being the displacement of the mirror,

\begin{eqnarray}
L_1&=&2(L_0+d_1)\\
L_2&=&2(2L_0+d_1+d_2)\nonumber\\
L_3&=&2(3L_0+d_1+d_2+d_3)\,\, \nonumber\\ 
...\nonumber
\end{eqnarray}
with 
\begin{eqnarray}
\label{eqn:dn1}
d_n &=& d(t-[(2n-1)\tau + \alpha_n ]) \quad \mbox{and}\\
\label{eqn:dn2}
\alpha_n &=& 2\sum\limits_{l=1}^{n-1}\frac{d_l}{c}-\frac{d_n}{c}
\end{eqnarray}

where $\tau=L_0/c$.
With the round trip length $L=2L_0$ and with $X=r_1r_2 e^{-ikL}$ we obtain



\begin{eqnarray}
E_{tot}=tE(1+Xe^{-2ikd_1} +X^2e^{-2ik(d_1+d_2)}\nonumber\\
+X^3e^{-2ik(d_1+d_2+d_3)} \cdots )\nonumber
\end{eqnarray}

Since by definition the optical spring $K_{OS}$ is the linear term in the expansion $F=F_0+ K_{OS} d + O(d^2)$, we now expand the exponential in $d_n$ and we group $d_n$ terms 

\begin{eqnarray}
E_{tot}&=&\frac{tE}{1-X}(1-2ikd_1 X-2ikd_2 X^2-2ikd_3 X^3+\cdots) \nonumber
\end{eqnarray}
Given that any correction from $\alpha_n$ [Eq. (\ref{eqn:dn2})] is quadratic in $d(t)$, we can again neglect it by definition, and find for the harmonic mirror motion (i.e. in the Fourier domain)
\begin{eqnarray}
d_n&=&x_0e^{i\Omega(t-(2n-1)\tau)}=x_0e^{i\Omega t}e^{-i\Omega(2n-1)\tau}\nonumber\\
&=&x_0e^{i\Omega t} \frac{Y^{2n}}{Y}\frac{Y}{Y}=Y^{2n-2}d_1
\end{eqnarray}

where $Y=e^{-i\Omega\tau}$. Thus we can write

\begin{eqnarray}
E_{tot}&=&\frac{tE}{1-X}\left [1-\frac{2ikd_1 X}{1-Y^2X}\right ]
\end{eqnarray}

where $d_1$ is a complex number. Since we have to take its real part $Re (d_k)=\frac{d_k+\bar{d}_k}{2}$,
we consider the field inside the cavity with $\bar{d}_k$ conjugate of $d_k$ and we obtain as total field:



\begin{eqnarray*}
E_{tot}=tE\left [\frac{1}{1-X}- \frac{2ikX}{2(1-X)}   \left ( \frac{d_1}{1-Y^2 X} +\frac{\bar{d}_1}{1-\overline{Y}^2 X}\right )\right]
\end{eqnarray*}

%
 
Using the following expression
 
\begin{eqnarray}
d_1=x_0e^{i\Omega(t-\tau)}=x_0e^{i\Omega t}e^{-i\Omega\tau}=xY  
\end{eqnarray}
 
we can now obtain the intracavity power expression by multiplying $E_{tot}$ by its conjugate
and considering only the linear terms of $x$ and $\bar{x}$

\begin{eqnarray}
P&=&E_{tot}\cdot \overline{E}_{tot}=-P_0t^2 [ \frac{ikY}{(1-\overline{X})(1-X)}\nonumber\\ 
&\times &\left( \frac{X}{1-Y^2 X}-\frac{\overline{X}}{1-Y^2\overline{X}} \right) x + cc ]
\end{eqnarray}

where we have also neglected the first constant term.

Once we have calculated the power we can obtain the radiation pressure force on the end mirror by $F_{rad}=\frac{2 r_2^2}{c}P$. Furthermore
we can also notice the similarity of the expression with the elastic force. Thus we recall that
in frequency domain and complex notation $K$ is defined by $F=-Kx$, the real form is thus

\begin{eqnarray*}
F'=Re[F]=-\frac{1}{2}(Kx+\overline{K}\bar{x})=-\frac{1}{2}(Kx+cc)
\end{eqnarray*}

Taking into account that we are calculating the radiation pressure on the end mirror, we need to consider an extra delay factor $Y$
for the calculation of the power which appears in the expression of $K$. The complex spring is then given by 
\newpage
\begin{eqnarray*}
K_{OS}=\frac{2 r_2^2}{c} P_0 t^2  \frac{2ikY^2}{(1-\overline{X})(1-X)} \left( \frac{X}{1-Y^2 X}-\frac{\overline{X}}{1-Y^2\overline{X}} \right) 
\end{eqnarray*}
which can be rewritten in the form of Eqs. (\ref{KOS_full_2}) and (\ref{eqn:K0}).

\subsection{Detuning}
Given the frequency detuning is $\delta=\omega_0-\omega_{res}$ and $\Omega=\omega-\omega_0$,
where $\omega_0$ is the carrier (subcarrier) frequency and $\omega_{res}=2\pi n\cdot c/L$ is the resonant frequency, we get the following expression:

%

\begin{eqnarray}
e^{-ikL}=e^{-i\delta 2\tau}
\end{eqnarray}


If we now replace $X$ and $Y$ we obtain the exact expression for $K_{OS}$:

\begin{eqnarray}
K_{OS}=&-P_0 t^2 r_2^2 \frac{4ike^{-2i\Omega\tau}}{c(1-r_1\!r_2e^{i2\delta\tau})(1-r_1\!r_2e^{-i2\delta\tau})}\times\nonumber\\
 & \left( \frac{r_1\!r_2e^{-i\delta \tau}}{1\!-\!r_1\!r_2e^{-2i\Omega\tau} e^{-i2\delta\tau}}
 \!-\!\frac{r_1\!r_2e^{i2\delta\tau}}{1\!-\!r_1\!r_2e^{-2i\Omega\tau}e^{i2\delta\tau}} \right) 
\end{eqnarray}

\subsection{Comparison}

To compare to existing literature we now expand the exponentials to linear order 
in $\Omega$ and $\delta$, 
$e^{-i\delta 2\tau}\approx 1-i\delta 2\tau$
and $e^{-i2\Omega \tau}\approx 1-i2\Omega \tau$:

\begin{eqnarray}
K_{OS} =& - P_0 t^2 r_2^2 \nonumber \\ 
 \times  &\frac{4ik(1-2i\Omega\tau)r_1r_2}{c(1-r_1r_2+r_1r_2i2\delta\tau)(1-r_1r_2-r_1r_2i2\delta\tau)}\nonumber\\
 \times  &\left[\frac{1-i2\delta\tau}{1-r_1r_2(1-2i\Omega\tau-i2\delta\tau)} \nonumber
- \frac{1+i2\delta\tau}{1-r_1r_2(1-2i\Omega\tau+i2\delta\tau)} \right] \\
\end{eqnarray}


We further simplify this equation using expressions for the $Finesse \approx \pi \frac{r_1r_2}{1-r_1r_2}= \pi FSR/\gamma$ and the free spectral range $FSR=1/2\tau$, introducing  the cavity bandwidth $\gamma$. We also neglect the $i\Omega\tau$, $i\delta\tau$ terms in the numerator since they correspond to a simple time delay. We obtain:

\begin{eqnarray}
K_{OS} & \approx & P_0 t^2 r_2^2 \frac{8k r_1r_2}{c(1-r_1r_2)^3}\frac{ \frac{\delta}{\gamma}}{(1+\frac{\delta^2}{\gamma^2})} 
\left[\frac{1}{1+\frac{\delta^2}{\gamma^2}-\frac{\Omega^2}{\gamma^2}+i2\frac{\Omega}{\gamma} }\right]\nonumber\\
\end{eqnarray}

which is equivalent to the expression already existing in the literature \cite{Barginsky02,Corbitt07}.

\subsection{Overcoupled cavity}

In the particular case of perfectly overcoupled cavity ($r_2=1$) $Finesse/\pi=2/T_1$ and $(1-r_1r_2)^2=T_1^2/2$ and the optical spring constant becomes

\begin{eqnarray}
K_{OS} & \approx & 128 P_0  \frac{\pi}{c\lambda T_1^2}\frac{ \frac{\delta}{\gamma}}{(1+\frac{\delta^2}{\gamma^2})} 
\left[\frac{1}{1+\frac{\delta^2}{\gamma^2}-\frac{\Omega^2}{\gamma^2}+i2\frac{\Omega}{\gamma} }\right]\nonumber\\
\label{eqn:overcoupled}
\end{eqnarray}

\subsection{Matched cavity}

In this case of a matched cavity ($r_1=r_2$) $Finesse/\pi=1/T_1$ and $(1-r_1r_2)^2=T_1^2$ and the optical spring constant remains the same as in Eq. (\ref{eqn:overcoupled}) except for the the factor 128 which has to be replaced with 16.

\section{TORSION PENDULUM MECHANICAL PLANT}
\label{app:B} 

Here we transform
the basis of coordinates $\{x_G,\Theta\}$  formed by the position of the center of gravity $x_G$ of the mirror and its rotation angle $\Theta$  with respect to the vertical axis passing from $x_G$ into a basis $\{x_A,x_B\}$ formed by the length of the cavities relative to beam $A$ and beam $B$ respectively. Thus the longitudinal and angular control of the mirror can be treated as the longitudinal control of the two above mentioned cavities. The basis can be expressed as
%

\begin{equation}
\label{eqn:BDEF}
\begin{pmatrix}
x_A \\ x_B
\end{pmatrix}
=
 \begin{pmatrix}
1& r_A\\1& r_B
\end{pmatrix} 
\begin{pmatrix}
x_G\\ \Theta
\end{pmatrix}
=
\mathcal{B}
\begin{pmatrix}
x_G\\ \Theta
\end{pmatrix}
\end{equation}

where $r_A$ and $r_B$ are the lever arms of the two beams with respect to $x_G$.

The equation of motion for the mirror is
\begin{equation}
\label{eqn:motion_matrix}
-\omega^2
\begin{pmatrix}
m &  \\ & I
\end{pmatrix}
 \begin{pmatrix}
x_G\\ \Theta
\end{pmatrix}
= 
\begin{pmatrix}
F_{tot}\\ T_{tot}
\end{pmatrix}
\end{equation}
where $I$ is the moment of inertia of the mirror of mass $m$. We now express the total force and the total torque exerted on the mirror
as function of the individual forces $F_A$ and $F_B$:

\begin{equation}
\label{eqn:FtotTtot}
\begin{pmatrix}
F_{tot} \\ T_{tot}
\end{pmatrix}
=
 \begin{pmatrix}
1& 1\\r_A& r_B
\end{pmatrix} 
\begin{pmatrix}
F_A\\ F_B
\end{pmatrix}
=
\mathcal{B}^{T}
\begin{pmatrix}
F_A\\ F_B
\end{pmatrix}
\end{equation}

Using Eqs. (\ref{eqn:FtotTtot}) and (\ref{eqn:BDEF}) in Eq. (\ref{eqn:motion_matrix}) we obtain the equation of motion in the ${x_A,x_B}$ basis:
\begin{equation}
\label{eqn:M}
-\omega^2
\left[
\mathcal{B}^{T-1}
\begin{pmatrix}
m &  \\ & I
\end{pmatrix}
\mathcal{B}^{-1}
\right ]
 \begin{pmatrix}
x_A\\ x_B
\end{pmatrix} 
=
\begin{pmatrix}
F_{A}\\ F_{B}
\end{pmatrix}
\end{equation}

\section{STABILITY IN TWO DIMENSIONS}
\label{app:C}
The control loop stability in multiple dimensions can be evaluated by considering the one-dimensional open-loop transfer function of every control filter (i.e. optical spring) while all other loops stay closed. Here we calculate these open-loop transfer functions for the two-dimensional case.

Referring to Fig.\,\ref{fig:block_loops}, we inject a signal 
$F_{\rm ext}$ into the path of beam $A$. The output of path A is $F_A$. 
Simultaneously we close the control loop relative to beam $B$
by feeding back the force $F_B$, which represents the output of path B.

We obtain the following expression:
\begin{equation}
HM
\left( \begin{array}{c}
0\\ -F_B
\end{array} \right)
+
HM
\left( \begin{array}{c}
F_{ext}\\0
\end{array} \right)
=
\left( \begin{array}{c}
F_A\\F_B
\end{array} \right)
\end{equation}
If we introduce the $2\times2$ matrix $S$,
\begin{equation}
S_A=
\left( \begin{array}{cc}
0 & 0\\
0 & 1
\end{array} \right)
\end{equation}
we can  write
\begin{equation}
HMS_A
\left( \begin{array}{c}
- F_A\\ - F_B
\end{array} \right)
+
HM
\left( \begin{array}{c}
F_{ext}\\0
\end{array} \right)
=
\left( \begin{array}{c}
F_A\\F_B
\end{array} \right)
\end{equation}
Using the vector $e_A^{T}=(1,0)$ we are able to extract the following open loop
transfer function related to cavity A:
\begin{equation}
OL_{A}=\frac{F_{A}}{F_{ext}}=e_A^{T}(\mathds{1}+HMS_A)^{-1}HMe_A
\end{equation}

The same open loop transfer function can be obtained considering an external signal injected into the loop of the beam $B$ while the loop of beam $A$ remains closed,
\begin{equation}
OL_{B}=\frac{F_{B}}{F_{ext}}=e_B^{T}(\mathds{1}+HMS_B)^{-1}HMe_B
\end{equation}

with $e_B^{T}=(0,1)$ and 

\begin{equation}
S_B=
\left( \begin{array}{cc}
1 & 0\\
0 & 0
\end{array} \right).
\end{equation}